\documentclass[]{spie}  %>>> use for US letter paper

\usepackage{amsmath,amsfonts,amssymb}
\usepackage{graphicx}
\usepackage[colorlinks=true, allcolors=blue]{hyperref}
\usepackage[per-mode = symbol]{siunitx}
\title{A low-loss, 24-mode laser-written universal photonic processor in a glass-based platform}

\author[a]{Andrea Barzaghi}
\author[a]{Maëlle Bénéfice}
\author[b,a]{Francesco Ceccarelli}
\author[a,b]{Giacomo Corrielli}
\author[a]{Valerio Galli}
\author[a]{Marco Gardina}
\author[a]{Vittorio Grimaldi}
\author[a]{Jakub Kaczorowski}
\author[a]{Francesco Malaspina}
\author[b,a]{Roberto Osellame}
\author[a]{Ciro Pentangelo}
\author[a]{Andrea Rocchetto}
\author[c]{Alessandro Rudi}

\affil[a]{Ephos, Inc., 333 Bush Street, San Francisco, CA 94104, USA}
\affil[b]{Istituto di Fotonica e Nanotecnologie - Consiglio Nazionale delle Ricerche, P.za Leonardo Da Vinci 32, 20133 Milan, Italy}
\affil[c]{SDA Bocconi School of Management, Via Roberto Sarfatti 10, 20136 Milan, Italy}

\authorinfo{Send correspondence to Roberto Osellame\\E-mail: roberto@ephos.io.\\Authors are in alphabetical order.}

\begin{document} 
\maketitle

\begin{abstract}
We report the fabrication of the first 24-mode universal photonic processor (UPP) realized through femtosecond laser writing (FLW), marking the most complex UPP demonstrated to date. Optimized for quantum dot emission at \SI{925}{\nano\metre}, the device exhibits total insertion losses averaging only \SI{4.35}{\decibel}, enabling its direct application in advanced multi-photon quantum experiments. Leveraging the versatility of FLW, we introduce suspended waveguides and precisely engineered 2D and 3D microstructures, significantly enhancing thermal isolation and minimizing power dissipation. As a result, our processor operates efficiently at less than \SI{10}{\watt}, requiring only a simple thermo-electric cooler for stable thermal management. The device exhibits exceptional performance after calibration, implementing Haar-random unitary transformations with an amplitude fidelity of \SI{99.7}{\percent}. This work establishes FLW-based integrated photonics as a scalable and robust platform for advancing quantum computing, communication, and sensing technologies.

\end{abstract}


% Introduction
\pagestyle{plain}
\setcounter{page}{1}

\section{INTRODUCTION}
\label{sec:intro}

Integrated photonics is crucial for quantum information processing and signal routing, providing a pathway to large-scale photonic devices \cite{pelucchi22, bogaerts20}. Contrary to bulk optical setups, photonic integrated circuits (PICs) offer high interferometric stability in a compact footprint and can be reconfigured via electrically driven phase shifters.

Universal photonic processors (UPPs) \cite{harris18, taballione22} — programmable PICs performing arbitrary unitary transformations — have recently attracted considerable interest. Examples include triangular \cite{reck94} and rectangular \cite{clements16} meshes of Mach–Zehnder interferometers (MZIs), demonstrated on multiple platforms such as silicon nitride \cite{taballione22, degoede22}, silica-on-silicon \cite{carolan15}, and femtosecond laser writing (FLW) \cite{dyakonov18, pentangelo22}. FLW exploits irradiation with ultrashort laser pulses of transparent dielectric materials to inscribe waveguides with low losses (below \SI{0.3}{\decibel\per\centi\metre}\cite{corrielli21}) and negligible birefringence \cite{corrielli18}, supporting complex 3D layouts \cite{crespi16, hoch22}.

Thermal reconfigurability relies on microheaters that locally heat the substrate, inducing a refractive index variation without adding photon losses. However, heat diffusion can cause thermal crosstalk with neighboring waveguides to the actuated one, thus introducing unwanted phase shifts in the photonic circuit; deep isolation trenches in correspondence to the thermal shifters can be implemented to mitigate this effect \cite{ceccarelli20}. Operation in vacuum further reduces crosstalk, but slows down the phase shifters' response \cite{ceccarelli20}.

Scaling UPPs to a larger number of optical modes requires quadratically more thermal phase shifters, which single-metal-layer designs struggle to accommodate \cite{ceccarelli20}. Simultaneously, it is crucial to maintain both low waveguide losses and total circuit length.

We address these challenges by fabricating more compact and curved isolation trenches that allow for the elimination of straight waveguide sections between directional couplers, significantly shrinking the total circuit length. We also employ a two-metal lithography process, using a high-resistance material for heaters and a low-resistance material for interconnections, thanks to the use of a dry photoresist that uniformly covers the microstructured substrate. These strategies yield denser, lower-loss, and more thermally efficient UPPs in the FLW platform.

% Circuit presentation
\section{24-MODE UNIVERSAL PHOTONIC PROCESSOR}
\label{sec:circuit}

This section presents the 24-mode universal photonic circuit fabrication process, from the bare glass substrate to the final prototype. The optical circuit consists in a mesh of MZIs, each one controlled by two thermal shifters.

\subsection{Optical circuit}
\label{subsec:optical}

% \begin{figure}[t]
%     \centering
%     \includegraphics[width=\textwidth]{img/layout.pdf}
%     \caption{Scheme of the 24-mode circuit.}
%     \label{fig:circuit_layout}
% \end{figure}

The fabrication process starts with the waveguide writing of the 24-mode photonic circuit in Corning EAGLE XG alumino-borosilicate glass, at a depth $d =$~\SI{35}{\micro\metre} from the glass surface. The waveguides are optimized for single-mode operation at a wavelength of $\lambda =$~\SI{925}{\nano\metre} to match the emission wavelength of widely used single photon sources based on InGaAs quantum dots. The distance between adjacent waveguides at the input/output of the processor is set at $p =$~\SI{82}{\micro\metre} for coupling with reduced-pitch fiber arrays. The curved arms between each of the 552 directional couplers are \SI{1}{\milli\metre} long and the minimum bending radius of the waveguides is set at \SI{15}{\milli\metre}. With this geometry, the full device occupies a footprint of $15\times$\SI{134}{\milli\metre^2}. These dimensions are comparable to the state of the art for FLW circuits \cite{ceccarelli20, albieropentagelo22}.

After waveguide fabrication and a thermal annealing step, the isolation trenches \cite{ceccarelli20} are ablated on each side of every waveguide. The trenches are \SI{60}{\micro\metre} deep and \SI{1}{\milli\metre} long. The substrate is suspended upside down in water while the laser pulses are coming from the opposite surface of the sample. After the ablation process, the MZI arms are suspended in air like bridges and isolated on the sides from the surrounding glass.

\subsection{Integration}
\label{subsec:integration}

\begin{figure}[t]
    \centering
    \includegraphics[width=0.6\textwidth]{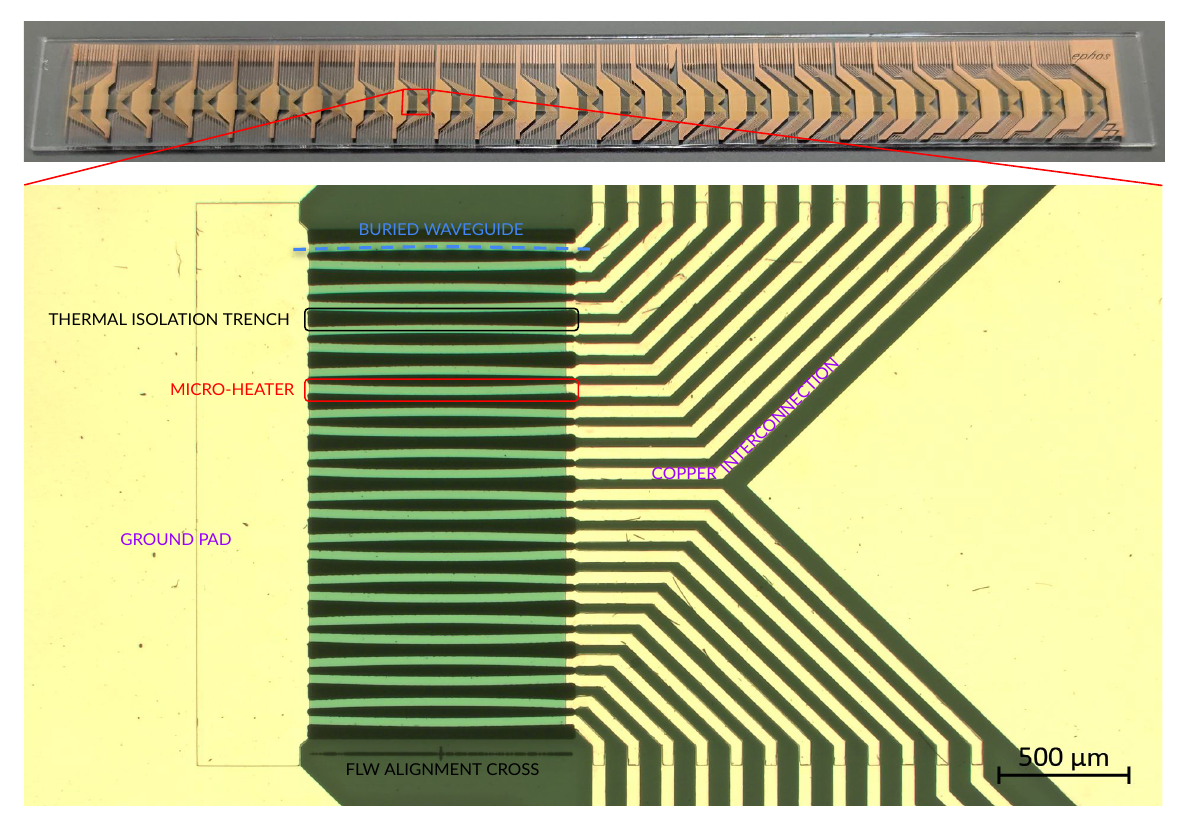}
    \caption{Picture of the 24-mode chip and microscope image of a single column of microheaters.}
    \label{fig:microheaters}
\end{figure}

The microheater fabrication \cite{albieropentagelo22} follows that of the trenches: the sample undergoes a proper cleaning of the top surface, where the \num{576} microheaters will be fabricated. A first maskless photolithography step defines the layout of the resistive chromium elements. A \SI{100}{\nano\metre} thin chromium layer is deposited by e-beam evaporation on top of the substrate and a lift-off procedure strips the remaining dry-photoresist. Thermal annealing is then performed in vacuum in order to stabilize the film for subsequent processing. A second evaporation encompasses the deposition of a thin titanium adhesion layer and a \SI{1}{\micro\metre}-thick copper layer that is selectively etched after a second photolithography step. The layout for the conductive interconnections is thus defined. A second annealing step ensures good contact between the chromium and copper layers, leading to a minimal parasitic resistance. The device is electrically characterized and then passivated with a silicon dioxide capping layer, thus finalizing the fabrication of the largest universal photonic processor in the literature, to our knowledge.

The chip is mounted on an aluminum sample-holder with thermally conductive glue and sandwiched between two lateral PCBs for electrical connectorization. Wire-bonding clinches electrical connections between copper pads on the chip and the PCBs. The photonic packaging encompasses the fiber pigtailing procedure that features low coupling losses due to the small mismatch between the standard fiber mode and the FLW waveguide mode.

% Results
\section{EXPERIMENTAL RESULTS}
\label{sec:results}

This sections presents the experimental results obtained on the 24-mode circuit, from the optical and electrical characterization to its calibration and performance evaluation.

\subsection{Device characterization}
\label{subsec:device_characterization}

\begin{figure}[t]
    \centering
    \includegraphics[width=\textwidth]{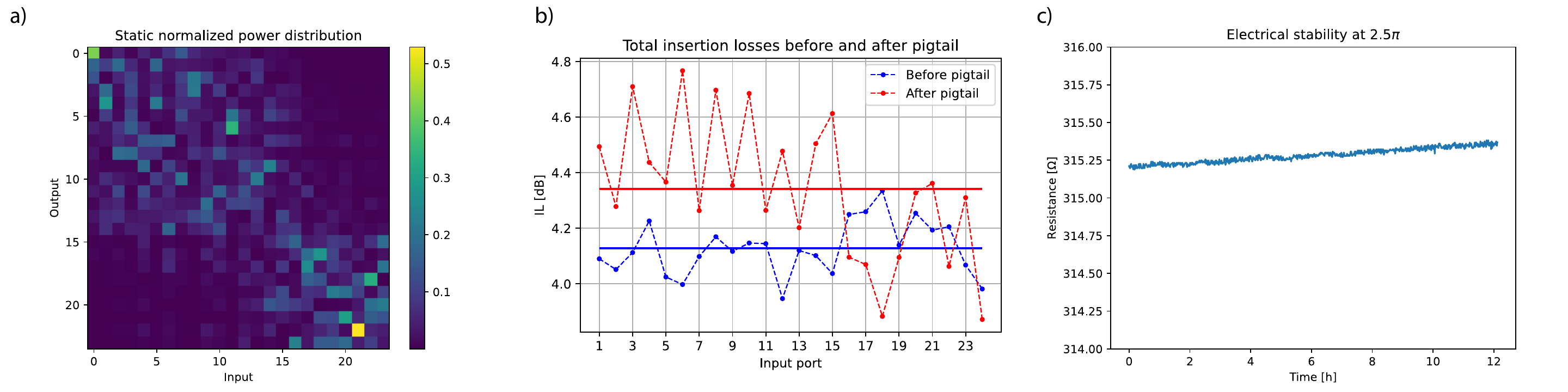}
    \caption{Static characterization of the processor. (a) Normalized power distribution for the static transformation implemented by the circuit, when all phase shifters are turned off. (b) Total insertion losses measured before and after pigtailing. (c) Electrical stability measurement performed on a typical phase shifter dissipating \SI{57}{\milli\watt} for 12 hours, corresponding to $2.5\pi$.}
    \label{fig:characterization}
\end{figure}

After fabrication, the circuit is characterized to assess its performance. First, \SI{925}{\nano\metre} laser light is coupled into each input mode and collected from all output modes both before and after pigtailing. From this measurement, we can construct the static optical transformation implemented by the circuit and evaluate the total insertion loss (IL) per input port (Fig.~\ref{fig:characterization}a,b). The average IL increases after pigtail from about \SI{4.1}{\decibel} to \SI{4.35}{\decibel}. This increase is higher than expected and may be due to the uneven distance between fiber cores in the fiber array used.

The phase shifters are characterized by measuring their resistance and stability over time. They are powered on to dissipate \SI{57}{\milli\watt} (corresponding to a phase shift of \num{2.5\pi}) for more than \num{12} hours to ensure that their resistivity does not change appreciably over this time span. In Fig.~\ref{fig:characterization}c, the measurement of a typical phase shifter is shown with an upward drift of \SI{<0.005}{\percent\per\hour}.

\subsection{Calibration}
\label{subsec:calibration}

\begin{figure}[t]
    \centering
    \includegraphics[width=\textwidth]{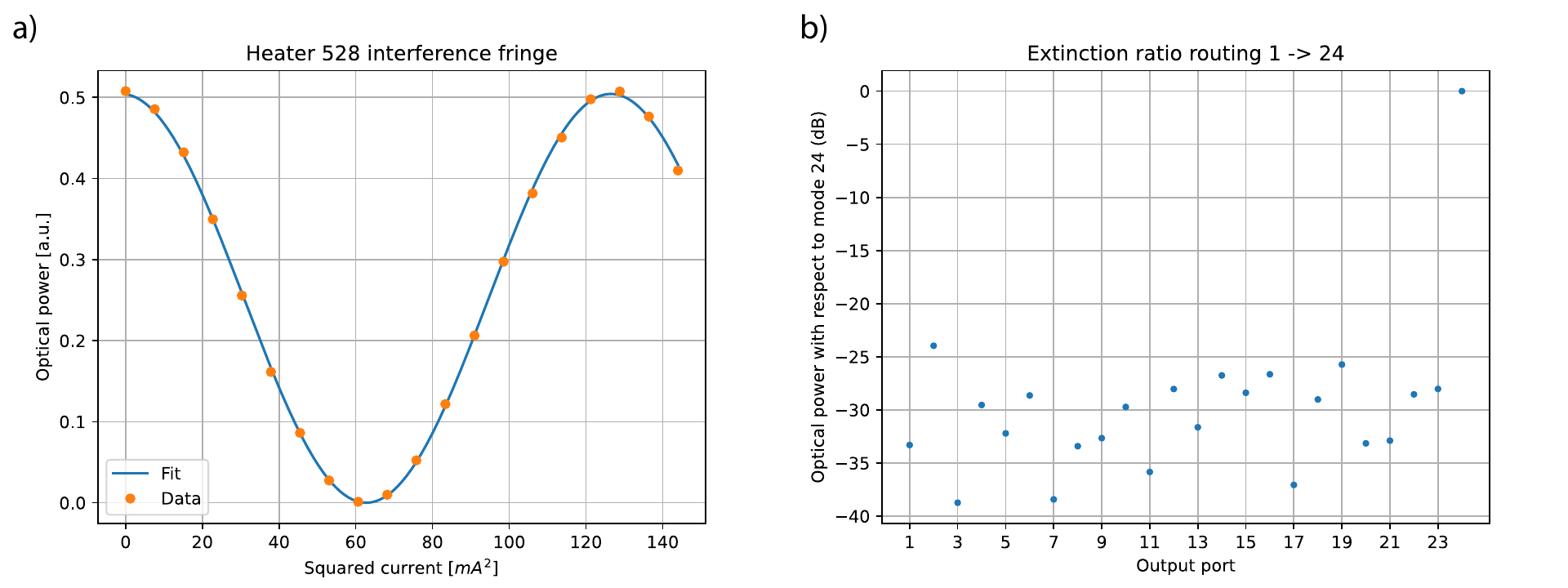}
    \caption{Routing 1-24 optimization. (a) Example of interference fringe measured on a phase shifter during the optimization of this routing. (b) Output optical power distribution with respect to mode 24 for the optimized routing.}
    \label{fig:routing}
\end{figure}

After the initial characterization, a calibration procedure is carried out. The final goal of the calibration is to fully program the circuit so that arbitrary unitary transformations can be implemented. The calibration starts by optimizing the routing from the first (top-most) input port to the last output port. During this optimization we can evaluate the electrical power required to dissipate a full \num{2\pi} phase shift on the shifters present along this route by performing interference fringe measurements. One of these measurements is shown in Fig.~\ref{fig:routing}a, where the power required for a \num{2\pi} shift is \SI{46}{\milli\watt}. The optical power distribution on all outputs for the optimized routing is shown in Fig.~\ref{fig:routing}b, where an extinction ratio of \SI{24}{\decibel} is achieved.

\begin{figure}[t]
    \centering
    \includegraphics[width=\textwidth]{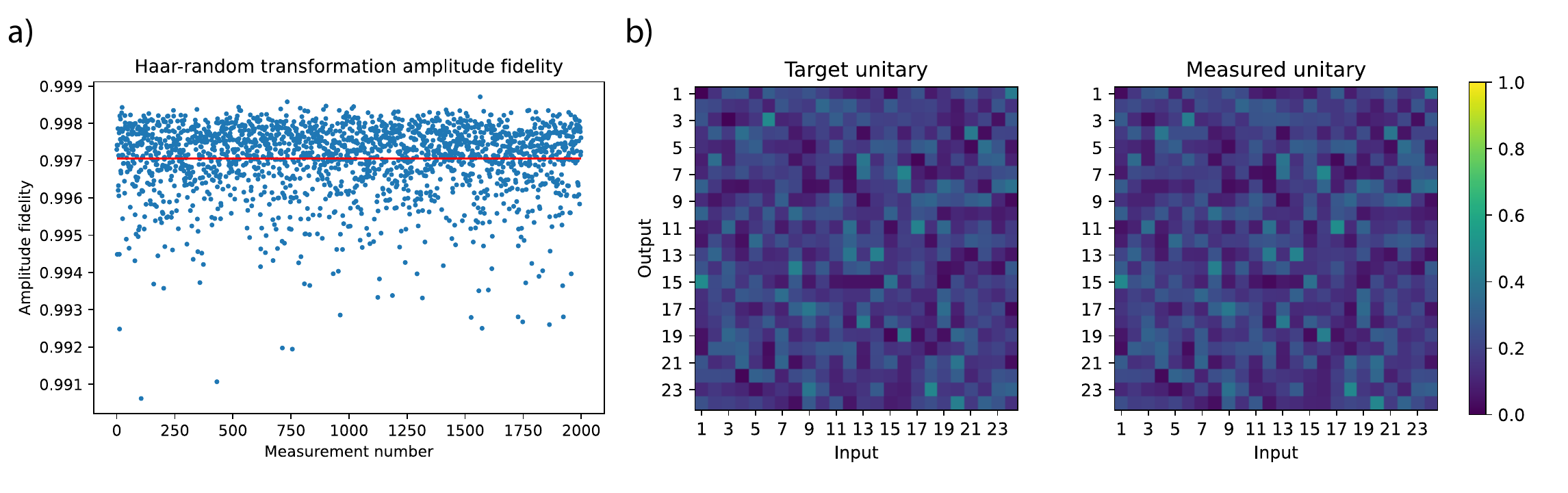}
    \caption{Amplitude fidelity for 2000 Haar random unitaries. (a) Scatter plot of amplitude fidelity for all measurements. The average (\SI{99.7}{\percent}) is highlighted by the red horizontal line. (b) Amplitudes of a target and measured Haar-random unitary transformation implemented with \SI{99.57}{\percent} amplitude fidelity.}
    \label{fig:haar_fids}
\end{figure}

To fully calibrate the circuit, a total of \num{30000} unitaries were measured, where all phase shifters were turned on with uniformly random dissipated power between \num{0} and \SI{45}{\milli\watt}. These unitaries were used to train a machine learning model of the circuit comprising a set of parameters including the \num{576} static phase contributions of each phase shifter, the \num{552} directional coupler splitting ratios and the \num{13824} thermal cross-talk coefficients. The calibration was tested by implementing 2000 Haar-random unitary matrices, achieving an average amplitude fidelity of \SI{99.7}{\percent}, as shown in Fig.~\ref{fig:haar_fids}a. Target matrices are well reproduced with this value of fidelity, as can be seen in the example reported in Fig.~\ref{fig:haar_fids}b. This performance represents the state-of-the-art for this type of processor\cite{taballione22}.

The total electrical power dissipated on the processor was lower than \SI{10}{\watt} for all arbitrary transformations tested; including Haar-random, switching and random phase.

% Conclusion
\section{CONCLUSION}
\label{sec:conclusion}

We have demonstrated a 24-mode FLW-UPP, representing the most complex device of its kind currently reported in the literature, to our knowledge. By introducing curved isolation trenches and employing a two-metal photolithography process, we substantially minimized crosstalk while maximizing thermal efficiency, thus enabling a denser circuit layout without sacrificing waveguide quality (\SI{4.35}{\decibel} fiber-to-fiber insertion losses). The bridge isolation structures, combined with an optimized design of conductive interconnections, allowed us to maintain a total electrical power consumption of less than \SI{10}{\watt} for all arbitrary unitaries implemented — well within manageable limits for standard thermo-electric cooling.

Our calibration procedure enables the operation of the processor with high accuracy, achieving an average amplitude fidelity of \SI{99.7}{\percent} on average for Haar-random unitary transformations. This level of precision attests the reproducibility and reliability of our FLW waveguides and micro-structures, as well as the robustness of the two-metal heater design. Altogether, our achievements emphasize the significant potential of femtosecond laser written universal photonic processors to drive forward the next generation of integrated quantum photonic circuits and optical signal processors.

\acknowledgments % equivalent to \section*{ACKNOWLEDGMENTS}       
This work is supported by the European Union’s Horizon Europe research and innovation program under QLASS (Quantum Glass-based Photonic Integrated Circuits, Grant Agreement No. 101135876) and FUTURE (Femtosecond laser writing for photonic quantum processors, Grant Agreement No. 101136471).

Fabrication of the resistive heaters for the femtosecond laser-written devices was performed at PoliFAB \cite{polifab}, the micro- and nano-fabrication facility of Politecnico di Milano. The authors would like to thank the PoliFAB staff for valuable technical support.

% References
\bibliography{report} % bibliography data in report.bib
\bibliographystyle{spiebib} % makes bibtex use spiebib.bst

\end{document}